# Robustness Metric for Quantifying Causal Model Confidence and Parameter Uncertainty

Garrett Waycaster[*], Christian Bes[†], Volodymyr Bilotkach[‡], Christian Gogu[§], Raphael T. Haftka[**], and Nam-Ho Kim[††]

**Abstract.** Many methods of estimating causal models do not provide estimates of confidence in the resulting model. In this work, a metric is proposed for validating the output of a causal model fit; the robustness of the model structure with resampled data. The metric is developed for time series causal models, but is also applicable to non-time series data. The proposed metric may be utilized regardless of the method selected for fitting the causal model. We find that with synthetically generated data, this metric is able to successfully identify the true data generating model in most cases. Additionally, the metric provides both a qualitative measure of model confidence represented by the robustness level as well as accurate estimates of uncertainty in model coefficients which are important in interpreting model results. The use of this metric is demonstrated on both numerically simulated data and a case study from existing causal model literature.

**1. Introduction.** Understanding relationships between multiple variables and predicting the effects of manipulations of variables is of key interest in many fields such as economics, health sciences, and engineering. Causal models are a tool that allows for estimation of the cause and effect relationships between variables that are not described by correlation based methods such as regression. Pearl [1] provides a thorough overview of the concepts and challenges of causality and identifying causal models. Causal models are generally represented by Bayesian networks, or graphically as a Directed Acyclic Graph (DAG) as shown in Figure 1 where $f$ represents a probability density function. In this case, conditional dependencies between variables in the network, represented by directed arrows in the DAG, can be interpreted as causal effects between variables.

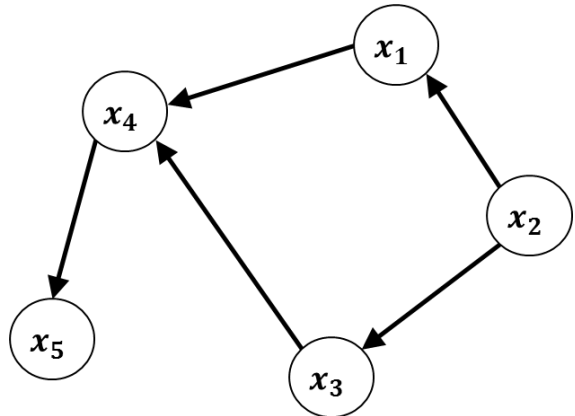

$$f(x_1, x_2, x, x_4, x_5) = f(x_5|x_4)f(x_4|x_1, x_3)f(x_1|x_2)f(x_3|x_2)f(x_2)$$

Figure 1. Directed acyclic graph for a Bayesian network. (From Chen and Chihying [2])

Causal effects are often estimated through use of experiments where causal hypotheses can be directly tested [1]. However in many cases, experimentally manipulating variables is not feasible; for these cases many methods have been developed for learning causal models based on observational data. Methods for fitting causal models

---
[*] Mechanical & Aerospace Engineering, University of Florida, Gainesville, Florida, 32611
[†] Institut Clément Ader, Université Toulouse III, Toulouse, 31400, FR
[‡] Newcastle University Business School, Newcastle University, Newcastle upon Tyne, NE1 8ST, UK
[§] Institut Clément Ader, Université Toulouse III, Toulouse, 31400, FR
[**] Mechanical & Aerospace Engineering, University of Florida, Gainesville, Florida, 32611
[††] Mechanical & Aerospace Engineering, University of Florida, Gainesville, Florida, 32611

include the Peter and Clark (PC) algorithm [3], the Spirtes-Glymour-Scheines (SGS) algorithm [4], the Inductive Causation (IC) algorithm [5], and the Sparsest Permutation (SP) algorithm [6] for non-temporal data and methods such as the Time Series Causal Model (TSCM) algorithm [2] for temporal data. Others such as Yoo [7] have developed methods for estimating causal models when some important variables are missing and identifying these variables, known as latent variables. Causal models may be used for many tasks, such as forecasting financial data [8] or interpreting the results of clinical trials for implementing new medicines or procedures [9].

Using any one of these methods, we may always estimate causal relationships between a given dataset and produce one (or sometimes several) causal models as output. However, it then falls to experts and analysts to determine if the resulting model is reasonable and should be trusted. The available data might be insufficient to make accurate estimates of causal effects or important variables may be missing, violating the common causal sufficiency assumption. While significant work has been devoted to selecting models based on minimizing unexplained variance in the observed data [10], these approaches may not necessarily result in correctly identifying model parameters. The objective of this work is to propose a metric to be used when fitting causal models which may assist in detecting when to accept a generated causal model. We find this metric also provides useful estimates of standard errors in model parameters in the case that the model is identified correctly. Understanding when to believe a fitted causal model and the uncertainty in its parameters is critically important for decision makers attempting to utilize these models. Though this work is focused on applications with time series data, the metric also may be simplified to non-time series models as well.

The outline of the paper is as follows: first, we introduce the basics of causal models and define our causal model metric, the robustness of a model structure with resampled noisy data. We then introduce several techniques for quantifying the accuracy of an estimated causal model as compared to a true data generating model. Using these techniques, we study the validity of the proposed metric with numerical simulated data. We then apply the metric to an existing case study from time series causal model literature.

**2. Introduction to causal models.** We will focus on the case of joint normally distributed data with linear causal effects, such that for multivariate time series data with $n$ variables and $T$ observations as shown in equation ( 2.1 ), we may represent a causal model as the vector autoregressive process for any arbitrary time $t$ as defined in equation ( 2.2 ).

$$( 2.1 ) \qquad x_t = \begin{bmatrix} x_{1t} \\ \vdots \\ x_{nt} \end{bmatrix} \; for \; t = 1, \dots, T$$

$$( 2.2 ) \qquad A_0^T x_t + A_1^T x_{t-1} + \dots + A_p^T x_{t-p} = \epsilon_t$$

where $A_0$ represents a matrix of the contemporaneous causal effects, $A_{p-k}$ represents a matrix of the temporal causal effects at time lag $k = 1, \dots, p-1$, respectively, and $\epsilon_t$ represents the residual error in the observed data at any arbitrary time $t$. Note that to avoid cycles in the model, there must exist a permutation of the matrix $A_0$ such that the matrix becomes triangular with ones along the main diagonal.

We may consider the matrix form of the causal model described in equation ( 2.2 ) as shown in equation ( 2.3 ).

$$( 2.3 ) \qquad A^T X_{t,p} = \begin{bmatrix} A_0^T & 0 & \cdots & 0 \\ A_1^T & A_0^T & \ddots & \vdots \\ \vdots & \ddots & \ddots & 0 \\ A_p^T & \cdots & A_1^T & A_0^T \end{bmatrix} X_{t,p} = \begin{pmatrix} \epsilon_{t-p} \\ \vdots \\ \epsilon_{t-1} \\ \epsilon_t \end{pmatrix} \; for \; p+1 \leq t \leq T$$

where $X_{t,p}$ is defined as a vector of the observed data as described in equation ( 2.4 ).

$$( 2.4 ) \qquad X_{t,p} = \begin{bmatrix} x_{t-p} \\ \vdots \\ x_t \end{bmatrix} \; for \; p+1 \leq t \leq T$$

Assuming we have a stationary process, the causal relationships for any vector $X_{t,p}, p+1 \leq t \leq T$ will be equivalent.

**3. Robustness metric for causal models.** The robustness metric is based on generating statistically similar datasets and refitting causal models to the newly generated data. By calculating the percentage of cases in which we observe the same model structure, we determine how robust each structure is to the inherent unexplained noise in the data.

The first step of being able to calculate model robustness is the ability to generate statistically similar data. Our ability to generate statistically similar data relies on several assumptions: that we have sufficient number of observations to be representative of the underlying process, that noise in the autoregressive process is uncorrelated, and that our time series is stationary. While these assumptions are significant, they are already required assumptions for accurate estimation of causal models and therefore generating new data does not impose additional restrictions.

For non-time series data we would commonly use bootstrapping, where we generate a new dataset by randomly resampling from our initial data with replacement. Since the ordering of observations is important for time series data, we utilize the autocovariance matrix of our initial dataset to generate new samples. The autocovariance matrix describes the relationships between variables at each time step as well as between different time steps as shown in equation ( 3.1 ).

$$( 3.1 ) \quad \Lambda = \begin{bmatrix} L(1,1) & L(1,2) & \cdots & L(1,k) \\ L(2,1) & \ddots & \ddots & \vdots \\ \vdots & \ddots & \ddots & L(k-1,k) \\ L(k,1) & \cdots & L(k,k-1) & L(k,k) \end{bmatrix} \; for\; k = 1, \dots, T$$

We may estimate the elements of the autocovariance matrix using equation ( 3.2 ) as the covariance of two time shifted sets $x_i$ and $x_j$ of our original $T$ observations, where for $n$ variables, each element $L$ will form an $n \times n$ matrix.

$$( 3.2 ) \quad L(i,j) = cov(x_i, x_j) \; for\; i = 1,\dots k, j = 1, \dots, k$$

Assuming we have a stationary time series process, we may utilize multiple samples of our time series with the same separation in time to estimate this covariance. Additionally, the autocovariance will not change over time and we may simplify the autocovariance as shown in equation ( 3.3 ).

$$( 3.3 ) \quad L(i,j) = L(i-j) = L(\tau) \equiv L_\tau, \tau = i - j;\; L(i,j) = L(j,i)^T$$

Using stationarity, we may therefore simplify the matrix in equation ( 3.1 ) as a symmetric block Toeplitz (diagonal-constant) matrix as shown in equation ( 3.4 ).

$$( 3.4 ) \quad \Lambda = \begin{bmatrix} L_0 & L_1 & \cdots & L_k \\ L_1^T & \ddots & \ddots & \vdots \\ \vdots & \ddots & \ddots & L_1 \\ L_k^T & \cdots & L_1^T & L_0 \end{bmatrix} \; for\; k = 1, \dots, T$$

Using this autocovariance matrix, we may calculate the distribution of our data as the current time step, $x_t$ conditional on the values of each $x$ at the previous $p$ time steps. The covariance of matrix of this conditional distribution is calculated using the Schur compliment of the autocovariance matrix as shown in equation ( 3.5 ).

$$( 3.5 ) \quad COV(x_t | x_{t-1,\dots,t-p}) = L_0 - [L_1 \;\; \cdots \;\; L_{p+1}] \begin{bmatrix} L_0 & L_1 & \cdots & L_p \\ L_1^T & \ddots & \ddots & \vdots \\ \vdots & \ddots & \ddots & L_1 \\ L_p^T & \cdots & L_1^T & L_0 \end{bmatrix}^{-1} \begin{bmatrix} L_1^T \\ \vdots \\ L_{p+1}^T \end{bmatrix} = \Gamma_0$$

Without loss of generality, we may consider each variable of our stationary time series to have zero mean. The conditional expectation for $x_t$ is then given as

$$( 3.6 ) \quad E(x_t | x_{t-1,\dots,t-p}) = [L_1 \;\; \cdots \;\; L_{p+1}] \begin{bmatrix} L_0 & L_1 & \cdots & L_p \\ L_1^T & \ddots & \ddots & \vdots \\ \vdots & \ddots & \ddots & L_1 \\ L_p^T & \cdots & L_1^T & L_0 \end{bmatrix}^{-1} \begin{bmatrix} x_{t-1} \\ \vdots \\ x_{t-p} \end{bmatrix} = W \begin{bmatrix} x_{t-1} \\ \vdots \\ x_{t-p} \end{bmatrix}$$

Combining equations ( 3.5 ) and ( 3.6 ), we may represent the distribution of $x$ at the current time as shown in equation ( 3.7 ).

$$( 3.7 ) \quad x_t | x_{t-1,\dots,t-p} \sim N\left(W \begin{bmatrix} x_{t-1} \\ \vdots \\ x_{t-p} \end{bmatrix}, \Gamma_0\right)$$

We may use this distribution to generate new statistically similar samples to our original data by sequentially generating samples. This requires a user specified initial condition for $x_{1,\ldots,p}$, we may choose to generate more samples than required and remove some number of initially generated samples such that the effect of this user specified initial condition is negligible on our final data. We also must specify some lag length $p$ in order to generate new data. This lag length may be chosen to be arbitrarily larger than the expected length of significant causal effects in order to ensure the validity of the newly generated data.

We may then utilize any potential time series causal model learning algorithm to estimate a causal model from many repetitions of this generated bootstrap data. We determine the robustness of a given model structure by calculating the percentage of these time series bootstrap samples which produced the same structure. Robustness may be calculated explicitly for $N$ runs with a single structure appearing $K$ times as shown in equation ( 3.8 ).

$$( 3.8 ) \quad R = 100\frac{K}{N}$$

We propose that the robustness value of a model structure found using any causal model fitting method is a useful tool to determine whether that structure is equivalent to the true data-generating model structure. We may also consider the value of maximum robustness as a measure of our confidence in the resulting model.

In addition to acting as a model validation criterion, the robustness metric may be used to approximate the uncertainty in the coefficients of the causal model. For every instance of the model structure in the bootstrap sampling, the estimate of model coefficients will be slightly different due bootstrapping producing different time series realization. We may therefore calculate the standard deviation of each model coefficient and use this as a measure of the standard error of the model coefficients, which can be critically important to analyzing the model in order to make decisions.

**4. Quantifying metric effectiveness.** In order to understand the efficacy of the proposed metric, we utilize several scoring criteria for causal model results when working with synthetically generated data with a known true coefficient matrix $A$. First, we consider whether the model structures suggested by the robustness metric is observationally equivalent to the true data-generating structure. Observational or Markov equivalence is defined by Pearl and Verma [5] as follows:

> "Two DAGs (models) are observationally equivalent if and only if they have the same skeleton and the same set of v-structures, that is two converging arrows whose tails are not connected by an arrow"

Put simply, observational equivalence means the chain of conditional dependencies does not represent a unique DAG, and therefore may not be exactly identified from observed data. Therefore, if a model that is observationally equivalent to the true model is returned by an algorithm, we consider this a success. Note that inspection of the model that is output by an algorithm will allow for identification of whether or not a model is observationally unique. For example, consider the causal model example from Figure 1. The relationships between $x_1$, $x_2$, and $x_3$ are not observationally unique as we could also have $x_1 \rightarrow x_2$ or $x_3 \rightarrow x_2$ (but not both) without creating or removing a v-structure, meaning the joint probability distribution described by these 3 possible models is equivalent.

In order to quantify the accuracy of the estimated coefficient values, we propose another scoring criterion utilizing the Frobenius norm of the difference of the true and fitted models normalized by the Frobenius norm of the true model. In order to construct this difference, we consider the full model coefficient matrix $A$ including contemporaneous and temporal effects as was utilized in equation ( 2.3 ). We then calculate this accuracy score as shown in equation ( 4.1 ).

$$( 4.1 ) \quad \zeta = \frac{\|A_{true} - A_{fit}\|_F}{\|A_{true}\|_F}$$

This results in the normalized root sum square error of the true versus predicted coefficient values. This score provides a quantitative measure of coefficient accuracy without requiring consideration of the model structure. This allows us to account for both when a model may have the correct structure but large error in coefficients, and also when a model might have a single incorrect structural element but be otherwise very close to the true model.

Finally, we propose a scoring criterion to better understand the ability to predict the uncertainty in model coefficient values. To do this, we compute the difference between true and estimated coefficients and divide this difference by the standard deviation of coefficient values obtained for the equivalent model structure by the robustness metric. This results in a matrix of errors in the coefficient estimates normalized by the estimated coefficient uncertainty. Recalling that the coefficient matrix $A$ will have dimension $n(p + 1) \times n(p + 1)$, where $n$ is the number of variables in $X$ and $p$ is the number of lags, this normalized error matrix $\Phi$ is calculated as

$$(4.2) \quad \Phi = [A_{true} - A_{fit}] \circ \begin{bmatrix} 1/\sigma_{\alpha_{1,1}} & \cdots & 1/\sigma_{\alpha_{1,n(p+1)}} \\ \vdots & \ddots & \vdots \\ 1/\sigma_{\alpha_{n(p+1),1}} & \cdots & 1/\sigma_{\alpha_{n(p+1),n(p+1)}} \end{bmatrix}$$

where $\circ$ represents the Hadamard product and $\sigma_{\alpha_{j,k}}$ is the standard deviation calculated for the $[j, k]$ non-zero element of the estimated coefficient matrix $A$ from the robustness metric. If this standard deviation is a reliable estimator of coefficient standard error, we may check that the elements of $\Phi$ where model coefficients are non-zero should follow a standard normal distribution.

**5. Application of robustness with the sparsest permutation algorithm.** For numerical testing, we utilize the sparsest permutation method developed by Raskutti and Uhler [6] to fit causal models. We find this method to be computationally when applied to time series data with small numbers of variables, which makes it well suited for numerical simulation with many repetitions. The description of this algorithm as presented in Rakutti and Uhler [6] is:

Let $S(G)$ denote the skeleton of a DAG $G$ and $|G|$ the number of edges in $G$ (or $S(G)$). Then the [sparsest permutation] algorithm is defined as follows:
(1) For all permutations $\pi$ of the vertices $\{1,2, \ldots, [n]\}$ construct [described below] $G_\pi$ and let $s_\pi = |G_\pi|$.
(2) Choose the set of permutations $\{\pi^*\}$ for which $s_{\pi^*}$ is minimal amongst all permutations.
(3) Output $G_{\pi^*}$ for all $\pi^*$ such that $s_{\pi^*}$ is minimal amongst all permutations.

The model $G$ is constructed by checking the conditional correlation between each variable. The edges of each DAG permutation are only retained when they satisfy the Markov condition; that is the conditional correlation of the connected variables given the preceding variables in the permutation is statistically non-zero. Therefore a directed edge between any two variables $j$ and $k$ for $j < k$ is retained if and only if the non-independence condition in equation ( 5.1 ) is satisfied at some user specified confidence level $\alpha$.

$$(5.1) \quad X_{\pi(j)} \not\perp X_{\pi(k)} | X_S \ where \ S = \{\pi(1), \pi(2), \ldots, \pi(k-1)\} \setminus \{\pi(j)\}$$

Raskutti and Uhler [6] show that for the case of joint normally distributed variables for non-time series data, this process can be simplified to the Cholesky decomposition of the inverse covariance matrix of our observations. In order to apply the sparsest permutation algorithm to time series data, we require some additional modifications to the method. The details of the basic sparsest permutation algorithm and our modifications are described in more detail in the Appendix.

To test our proposed metric, we randomly generate causal model structures $A$ and noise levels described by the diagonal matrix $D_0$ as shown in equation.

$$(5.2) \quad D_0 = \begin{bmatrix} \sigma_1^2 & 0 & 0 \\ 0 & \ddots & 0 \\ 0 & 0 & \sigma_n^2 \end{bmatrix} = cov(\epsilon)$$

We then synthesize observed data for the generated model, apply the proposed metrics, and compare the resulting fitted model to the known true model by the scoring criteria proposed in the previous section. Once the $A$ and $D_0$ matrices are created, we may generate any number of synthetic time series observations by drawing from the joint normal distribution specified by equation ( 5.3 ). This relationship is found directly by solving equation ( 2.3 ) for $X_{t,p}$.

$$(5.3) \quad X_{t,p} \sim N(0, (AD^{-1}A^T)^{-1})$$

Coefficients are selected randomly from the interval $[-1, -0.4] \cup [0.4, 1]$ such that the coefficients are bounded away from zero. In order to avoid fully connected models which will not be observationally distinguishable, we randomly set coefficients in $A$ to zero until meeting a target connectivity ratio $r$ defined as

$$(5.4) \quad r = \frac{m}{\frac{n(n-1)}{2} + n^2 p}$$

where $m$ is the number of non-zero coefficients in the generated model $A$ and $\frac{n(n-1)}{2} + n^2 p$ is the maximum number of coefficients in an $n$ variable, $p$ lag model.

Noise values are selected from a lognormal distribution such that the diagonal elements of $D_0$ are defined as shown in equation ( 5.5 ). This distribution results in [5%, 95%] bounds for these variances from [0.31, 0.43]. Note that increasing the level of noise does not have a significant effect on the sparsest permutation algorithm or the robustness metric so long as we have sufficient data to accurately estimate the autocovariance.

( 5.5 ) $$D_{ii} \sim logN(-1, 0.1)$$

The stationarity of the generated time series is ensured by checking that the roots of the characteristic polynomial of the vector autoregressive process as defined in equation ( 5.6 ) lie outside the unit circle such that

( 5.6 ) $$For\ any\ z, |z| \leq 1 \Rightarrow \det(I - A_0^{-1} A_1 z - \cdots - A_0^{-1} A_p z^p) \neq 0$$

Due to the computational expense of generating random models that meet this stationarity condition, we limit our numerical examples to time series of 3 variables with 1 or 2 lag periods. We then perform a case study applying the method of sparsest permutation for time series with the robustness selection metric for 1 or 2 causal lags, connectivity ratios of 0.4 or 0.5, and number of observations, $T$, set to be 500 or 1000. For each case, 1000 random causal models are generated and tested.

**6. Numerical results.** First, we consider the rate of success of the robustness metric to recover the true causal structure. Table 1 presents the percentage of the 1000 repetitions where the metric produced successful recovery of the model structure. It can be seen that the most robust model is the correct model in at least 78% of trials, with higher success rates with increasing sample size and smaller models. Across all model size cases, the sparsest permutation algorithm sees an 88% correct identification rate, which agrees well with the results of a similar study for non-time series data from Rakutti and Uhler [6]. Note that based on using 1000 repetitions, uncertainty in these percentages is roughly 1-2% at the 90% confidence level. This may explain the higher recovery rate for ($r = 0.4, p = 2$) as compared to ($r = 0.4, p = 1$) with 1000 observations. The recovery rate of the sparsest permutation with the original data, not considering robustness, is included in parentheses. Note that this rate includes cases where the sparsest permutation algorithm outputs multiple equally sparse, non-equivalent results and only one is correct. Conversely, the robustness metric can always provide a single most robust structure.

Table 1. Robustness method structure recovery rate for $p$ lags and connectivity ratio $r$ (recovery rate of sparsest permutation without considering robustness)

| | 500 observations | |
|---|---|---|
| | $r = 0.4$ | $r = 0.5$ |
| $p = 1$ | 95% (93%) | 88% (92%) |
| $p = 2$ | 87% (92%) | 78% (78%) |
| | **1000 observations** | |
| | $r = 0.4$ | $r = 0.5$ |
| $p = 1$ | 96% (95%) | 90% (90%) |
| $p = 2$ | 97% (91%) | 86% (84%) |

We may also consider how the calculated level of robustness is related to the chance of a model being identified correctly. If we find that low robustness models are more likely to be incorrect, we may choose only to trust models with a robustness greater than some threshold value. Figure 2 and Figure 3 show histograms of the robustness level of correctly and incorrectly identified models, respectively. These figures group all lag lengths, connectivity ratios, and sample sizes together as these factors are not found to influence the relationship between structure recovery and robustness level.

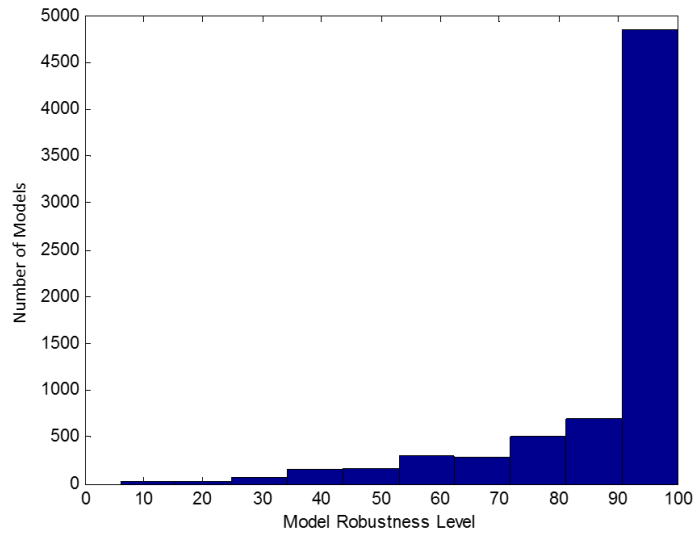

Figure 2. Number of correctly identified structures by robustness level (n = 7073)

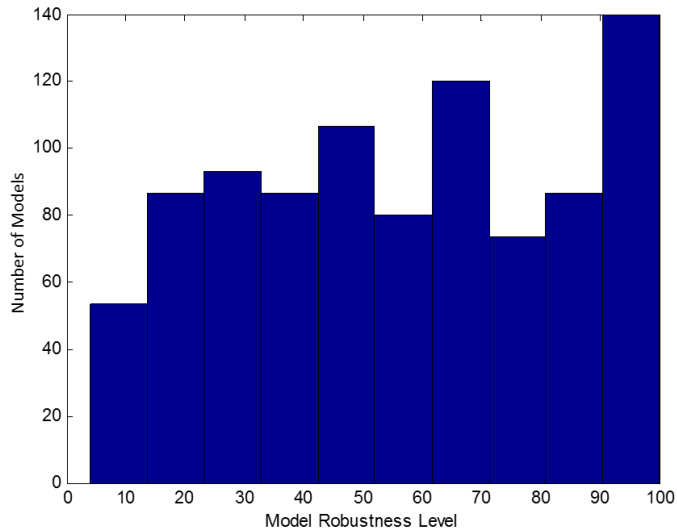

Figure 3. Number of incorrectly identified structures by robustness level (n = 927)

We find that correctly identified models typically have a robustness level above 70-80% while incorrectly identified structures are approximately uniform across all robustness levels. This indicates we could use a threshold robustness value below which we may elect not to trust the causal model output. For example, only considering cases with a robustness of 90 or more, the identification rate improves from 88% to 97%, while at a robustness of 55% or lower the chance of correctly identifying the true model falls below 50%. The robustness threshold could be specified by an analyst or expert based on the application and required model confidence.

Next, we consider our coefficient accuracy score based on the normalized Frobenius norm of the difference in fitted and true coefficient matrices. Figure 4 and Figure 5 present box plots of the accuracy metric $\zeta$ for the most robust model across the 1000 random models for each sample size, number of lags, and connectivity ratio for models selected by the robustness metric.

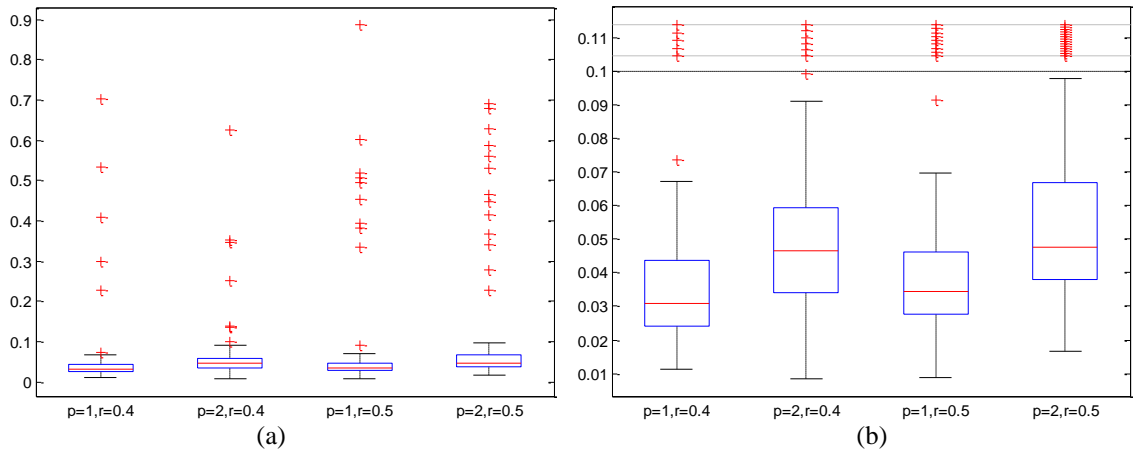

Figure 4. Accuracy score for each model size with 500 observations in (a) full and (b) zoomed plots

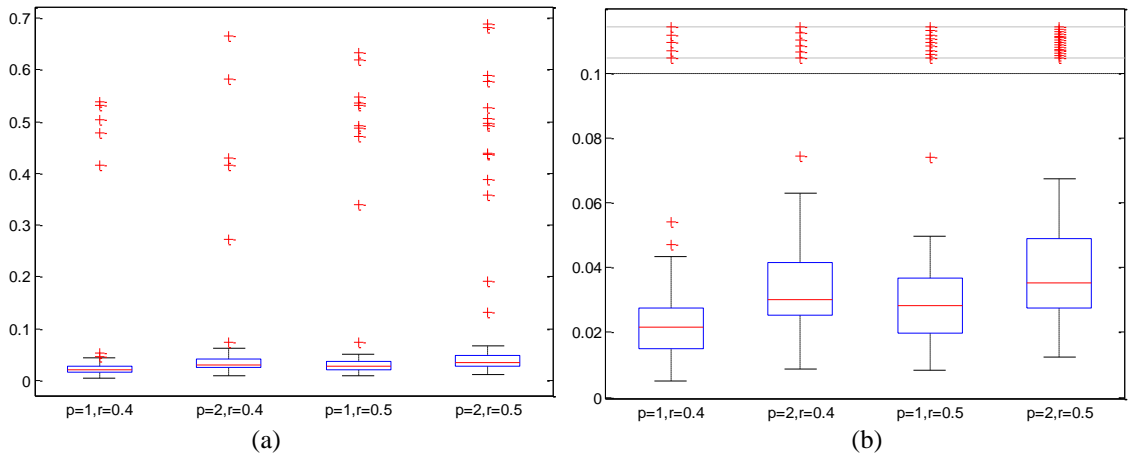

Figure 5. Accuracy score for each model size with 1000 observations in (a) full and (b) zoomed plots

We also consider the difference in accuracy score between the correctly and incorrectly identified models. We find this to be fairly consistent across different model sizes, so these results are all combined in Figure 6.

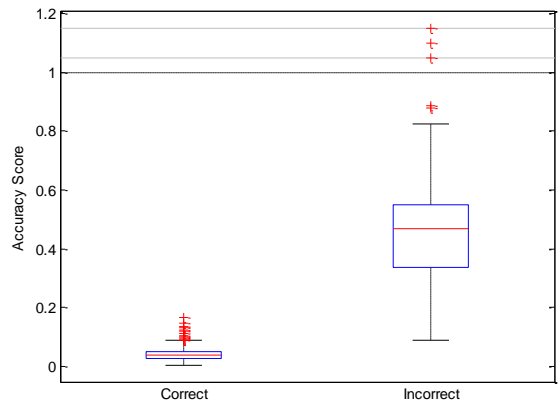

Figure 6. Accuracy score for correctly and incorrectly identified models

We find that the robustness metric is relatively consistent across all model sizes in terms of accuracy score. Accuracy score is improved slightly with sparser models, lower model lags, and increased sample size but in general will be less than 0.1. This indicates that the parameter values in the model coefficient matrix $A$ will be identified accurately by the most robust model. We may also consider the relationship between robustness level and accuracy score. This relationship for all model sizes and numbers of samples is presented in Figure 7.

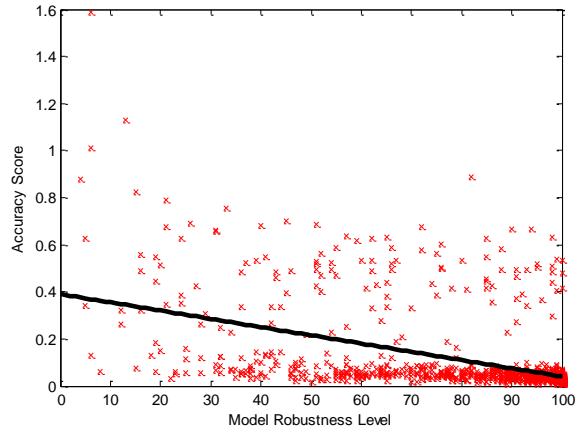

Figure 7. Comparison of model robustness and accuracy score for all cases

We observe a correlation between model robustness and accuracy score, with increasing robustness related to more accurate models. As with the structure recovery rate, this indicates that the model robustness level may be utilized as a diagnostic to determine the confidence in the accuracy of the estimated model parameters.

Finally, we calculate the error in coefficient estimation normalized by the standard deviation from the robustness metric. Because these results are consistent across each case considered, we compile all results together. Figure 8 presents a probability plot of the observed normalized coefficient errors for all models.

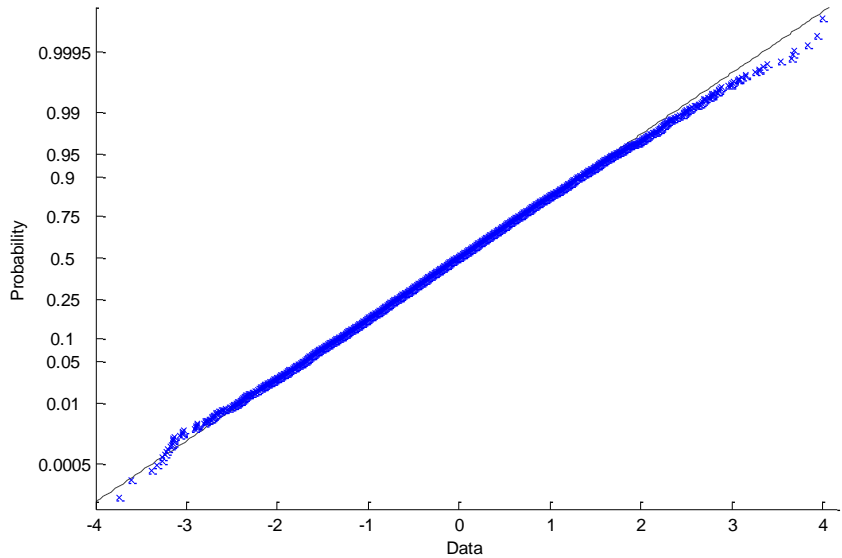

Figure 8. Probability plot of normalized error for all model sizes

We find that this normalized error does follow a standard normal distribution, suggesting that the coefficient standard deviation is a reasonable estimate for the error in estimation of causal model coefficients. This standard deviation may therefore be used to understand possible causal effects not otherwise available in causal model fitting methods.

**7. Wage-price dynamics case study.** In order to test our method with realistic data, we attempt to reproduce the findings of Chen and Chihying [2] regarding wage price dynamics using economic data from 1965-2004. We utilize the same dataset which is provided by the authors in their previous paper [11] and is drawn from the Federal Reserve economic database (FRED) [12]. The 6 variables used in the model defined by Chen and Chihying as $(w, p, e, u, z, \pi_m)$ are respectively wage inflation, price inflation, labor utilization rate, capacity utilization rate, growth of labor productivity, and inflationary climate. These variables are made stationary using the transformations described in Table 2.

Table 2. Raw data used for empirical investigation of the model [2]

| Variable | Transformation | Mnemonic | Description of the untransformed series |
|---|---|---|---|
| $e$ | log(1-UNRATE/100) | UNRATE | Unemployment Rate (%) |
| $u$ | log(GDPC1/GDPPOT) | GDPC1,GDPPOT | GDPC1: Real Gross Domestic Product of Billions of Chained 2000 Dollars, GDPPOT: Real Potential Gross Domestic Product of Billions of Chained 2000 Dollars, $u$: Capacity Utilization: Business Sector (%) |
| $w$ | log(HCOMPBS) | HCOMPBS | Business Sector: Compensation Per Hour, Index 1992=100 |
| $p$ | log(IDPBS) | IDPBS | Business Sector: Implicit Price Deflator, Index 1992=100 |
| $z$ | log(OPHPBS) | OPHPBS | Business Sector: Output Per Hour of All Persons, Index 1992=100 |
| $\pi_m$ | MA($dp$) | | Inflationary climate measured by the moving average of price inflation in the last 12 periods |

We attempt to fit models to this data using both the sparsest permutation algorithm used in our numerical study as well as the algorithm proposed by Chen and Chihying which we refer to as the time series causal model (TSCM) method. The TSCM method utilizes a two-step process to first optimize the contemporaneous effects, $A_0$, using a structural equation model approximation. Temporal effects are then optimized using a vector autoregression approximation, corrected for $A_0$ from the initial step. In both steps, the model structure is optimized using a greedy search algorithm which searches over all nearest neighbors of a random initial model structure by adding or removing one model connection. This search is repeated with multiple random initial structures. Models are scored for optimization using the Bayesian Information Criterion (BIC). More details of the algorithm may be found in the original paper [2].

The model found by Chen and Chihying is presented in Figure 9.

$$A_0 = \begin{pmatrix} 1 & 0 & 0 & 0 & 0 & 0 \\ 0 & 1 & 0 & 0 & 0 & 0 \\ 0 & 0 & 1 & 0 & 0 & 0 \\ -2.73 & 0.38 & -3.18 & 1 & 0 & 0 \\ -0.36 & 0 & 0 & 0.05 & 1 & 0 \\ 1.56 & -0.47 & 0 & -0.50 & -1.93 & 1 \end{pmatrix}$$

$$A_1 = \begin{pmatrix} -1.03 & 0 & 0 & 0 & 0.28 & 0 \\ -0.19 & -0.52 & -0.40 & 0.07 & 0 & 0 \\ 0.02 & -0.12 & -0.90 & 0 & -0.08 & 0 \\ 0.64 & 0 & 0 & 0 & 0 & -0.21 \\ 0.30 & 0.02 & 0 & 0 & -0.91 & 0 \\ -2.01 & 0 & -0.54 & 0 & 1.94 & 0 \end{pmatrix}$$

Figure 9. Chen and Chihying causal model using TSCM method

We find that we are able to reproduce this model using the TSCM method described in the original paper, while the sparsest permutation produces different results as shown in Figure 10. Specifically, the contemporaneous matrix $A_0$ has the permutation of the structure mostly reversed, while the temporal matrix $A_1$ more closely agrees with the TSCM result. This may be due in part to the fact that the sparsest permutation fits both $A_0$ and $A_1$ simultaneously rather than sequentially. Both methods also utilize very different scoring and search criteria which may lead to divergent results when the available data is insufficient. When adequate data is available in our numerical testing, we do find that both algorithms produce similar results. In order to reproduce model coefficients on the order of 0.02 as shown in the Chen and Chihying model, the sparsest permutation method must also use a very low confidence level when enforcing sparsity as described in equation ( 5.1 ); this has a significant effect on the model fit.

$$A_0 = \begin{pmatrix} 1 & -0.04 & -0.16 & -0.21 & -0.91 & 0.05 \\ 0 & 1 & 0 & 0 & 0 & -0.13 \\ 0 & -0.06 & 1 & 0 & 0.29 & 0 \\ 0 & 0.55 & 0 & 1 & -1.29 & -0.38 \\ 0 & 0 & 0 & 0 & 1 & 0 \\ 0 & 0 & 0 & 0 & 0 & 1 \end{pmatrix}$$

$$A_1 = \begin{pmatrix} -1.00 & 0 & 0.15 & 0 & 0.95 & 0 \\ -0.14 & -0.50 & -0.29 & 0.09 & 0 & -0.08 \\ 0 & -0.07 & -0.88 & 0 & -0.28 & 0 \\ 0.89 & -0.20 & 0.13 & 0.07 & 0.66 & -0.20 \\ -0.19 & 0.07 & 0.04 & 0 & -0.77 & 0 \\ -0.22 & -0.18 & -0.61 & 0 & -0.29 & 0 \end{pmatrix}$$

Figure 10. Sparsest permutation model results for wage-price dynamics

Applying the robustness metric for either the TSCM method or the sparsest permutation method with low confidence results in a model robustness of effectively zero; each bootstrap sample results in a different model structure. This implies that the available data is not sufficient to support the very small coefficient values included in the Chen and Chihying model. We can, however, increase the confidence level for enforcing sparsity until the robustness metric produces repeatable models. A similar approach may be taken with the TSCM method by modifying the penalty on model degrees of freedom in the BIC score, though this modification is less intuitive. Using a higher confidence level, the robustness metric applied to the sparsest permutation method produces the model shown in Figure 11 with a robustness of 29%.

$$A_0 = \begin{pmatrix} 1 & 0 & 0 & 0 & 0 & 0 \\ 0 & 1 & 0 & 0 & 0 & 0 \\ 0 & 0 & 1 & 0 & 0 & 0 \\ 0 & 0 & 0 & 1 & 0 & 0 \\ 0 & 0 & 0 & 0 & 1 & 0 \\ 0 & 0 & 0 & 0 & 0 & 1 \end{pmatrix}$$

$$A_1 = \begin{pmatrix} -1.19 & 0 & 0 & 0 & 0 & 0 \\ 0 & 0 & 0 & 0 & 0 & 0 \\ 0 & 0 & -0.87 & 0 & 0 & 0 \\ 0.80 & 0 & 0 & 0 & 0 & 0 \\ 0 & 0 & 0 & 0 & -0.90 & 0 \\ 0 & 0 & 0 & 0 & 0 & 0 \end{pmatrix}$$

Figure 11. Robustness metric suggested model for wage-price dynamics

This robust model suggests that only 4 coefficients from the original model are supported sufficiently by the available data. Considering that even this model has a robustness level of only 29%, one should still be hesitant to assume this model is correct. We may also consider the estimates of standard error in each coefficient obtained by the robustness metric, which are shown in Figure 12. Comparing the robust coefficient estimates and their standard error with the original Chen and Chihying model shows reasonable agreement. We may therefore consider that we have higher confidence in the values of these coefficients based on the available data. Overall, the high sparsity of the model compared to the Chen and Chihying model and low robustness suggest that additional data is likely needed in order to quantify some of the more complex effects governing these wage and price dynamics in question, depending on the level of confidence required by analysts. This finding agrees with previous literature such as Thompson [13] who discusses the difficulty of estimating covariance when using small sample sizes with data clustered both across time and across variables such as with the time series data considered here.

$$\sigma_{A_1} = \begin{pmatrix} 0.05 & 0 & 0 & 0 & 0 & 0 \\ 0 & 0 & 0 & 0 & 0 & 0 \\ 0 & 0 & 0.03 & 0 & 0 & 0 \\ 0.13 & 0 & 0 & 0 & 0 & 0 \\ 0 & 0 & 0 & 0 & 0.03 & 0 \\ 0 & 0 & 0 & 0 & 0 & 0 \end{pmatrix}$$

Figure 12. Robust model temporal coefficient standard error

**8. Conclusions.** In this work, we have described the robustness metric as a tool for quantifying confidence in causal models as well as uncertainty in model parameters, regardless of the method of fitting the causal model. We calculated robustness for time series data by generating statistically similar data using the autocovariance matrix and refitting the model many times. Several scoring criteria were introduced in order to quantify the effectiveness of this robustness metric. Though numerical simulations, we showed that using this metric: the most robust model often has the same structure as the true data generating model, the most robust model generally provides accurate estimates of model coefficients, the robustness level is a useful indicator of the level of confidence in the model structure and coefficients, and that the robustness metric provides accurate estimates for standard error in model coefficients. We applied this robustness metric to a case study problem from existing causal model literature and found that only a few of the estimated coefficients in the original model are robust to slight changes in the observed data. By understanding the confidence in fitted causal models and being able to estimate coefficient uncertainty, decision makers may be able to make more informed choices when attempting to work with causal models.

**ACKNOWLEDGEMENTS**

The authors gratefully acknowledge the support of the National Science Foundation grants CMMI-0927790 and 1131103 which funded this work.

# APPENDIX

**A. Sparsest permutation for time series derivation.** For the case of joint normal variables with linear causal effects for non-time series data, we can represent a DAG structure as a linear recursive simultaneous equation model (SEM),

(A.1) $$x_{jt} = \sum_{k=1}^{j-1} a_{jk} x_{kt} + \varepsilon_{jt} \; for \; j = 1, \ldots, n$$

or in equivalent matrix form,

(A.2) $$A_0^T x_t = \varepsilon$$

where $A_0$ is an upper triangular matrix with ones on the diagonal of the form

(A.3) $$A_0 = \begin{bmatrix} 1 & -a_{12} & \cdots & -a_{1n} \\ 0 & 1 & \ddots & \vdots \\ \vdots & \ddots & \ddots & -a_{(n-1)n} \\ 0 & \cdots & 0 & 1 \end{bmatrix}$$

Because of the assumption of joint normality, we know that $\varepsilon$ will be normally distributed and independent, such that

(A.4) $$A_0^T x_t \sim N(0, D_0)$$

where $D_0$ is the diagonal covariance matrix of $\varepsilon$.

We may solve this expression in equation (A.4) for $x_t$ as

(A.5) $$x_t \sim N(0, \Sigma) = N(0, (A_0 D_0^{-1} A_0^T)^{-1})$$

From this we may see that the covariance of the observed data, $\Sigma$, is enough to determine the matrix of causal model coefficients $A_0$ by using the Cholesky decomposition. By calculating the Cholesky decomposition of every permutation of $\Sigma$ and enforcing sparsity using equation (5.1), we may find the model (or set of models) which are the most sparse, meaning the most parsimonious representation of the causal model. Raskutti and Uhler show that this approach is valid under strictly weaker conditions than traditional score based model fitting approaches [6].

For the case of time series data, we are interested in inferring causal effects that may take place across several time periods. Under similar assumptions of normality and linear causal effects, these time series effects may be represented using a vector autoregression (VAR) model with a lag length $p$, representing the number of prior periods for which causal effects may exist. This idea was described previously by Chen and Chihying [2]. This VAR model may again be expressed as

(A.6) $$A_0^T x_t + A_1^T x_{t-1} + \cdots + A_p^T x_{t-p} = \epsilon_t$$

where $x_t$ is a vector of one observation of each variable at time $t$. Assuming we have a stable process such that these effects do not vary over time, we can again rewrite this in a matrix form for the complete time series as previously shown in equation (2.3).

(2.3) $$A^T X_{t,p} = \begin{bmatrix} A_0^T & 0 & \cdots & 0 \\ A_1^T & A_0^T & \ddots & \vdots \\ \vdots & \ddots & \ddots & 0 \\ A_p^T & \cdots & A_1^T & A_0^T \end{bmatrix} X_{t,p} = \begin{pmatrix} \epsilon_{t-p} \\ \vdots \\ \epsilon_{t-1} \\ \epsilon_t \end{pmatrix} \; for \; p + 1 \leq t \leq T$$

This large matrix on the left hand side of the equation containing the causal effects can be referred to as $A$, and similarly the matrix describing the variance of the independent errors $\epsilon$ will be a diagonal matrix $D$ where the diagonal elements are the variances of $\epsilon$, $D_0$, repeated $T$ times. We may therefore describe this time series similarly to the case of time independent observations as

(A.7) $$X_{t,p} \sim N(0, (AD^{-1}A^T)^{-1})$$

However, when estimating causal models from data, we often only have one observation for each variable at each time step. We are therefore unable to directly estimate the sample covariance matrix, $\hat{\Sigma} = (ADA^T)^{-1}$. However, we can determine the cross covariance, $\Lambda$, of a stationary time series up to $T - 1$ lags, and we can use this information to calculate the conditional distribution of the current time step data, $x_T$, given the data at the previous $p$ lags. This process is described in section 3, and the resulting conditional distribution was given in equation ( 3.7 ) shown again below.

( 3.7 ) $$x_t|x_{t-1,\ldots,t-p} \sim N\left(W\begin{bmatrix}x_{t-1}\\ \vdots \\ x_{t-p}\end{bmatrix}, \Gamma_0\right)$$

By subtracting the mean term, $Wx_{t-1,\ldots,t-p}$, from both sides and performing a Cholesky decomposition of $\Gamma_0$, we get

( A.8 ) $$x_t|x_{t-1,\ldots,t-p} - W\begin{bmatrix}x_{t-1}\\ \vdots \\ x_{t-p}\end{bmatrix} = N(0, (A_0 D_0^{-1} A_0^T)^{-1})$$

Multiplying both sides by $A_0^T$ results in the final expression,

( A.9 ) $$A_0^T x_t|x_{t-1,\ldots,t-p} - A_0^T W\begin{bmatrix}x_{t-1}\\ \vdots \\ x_{t-p}\end{bmatrix} = N(0, D_0) = \epsilon_t$$

where the time lagged causal effect matrices $A_1$ to $A_p$ are given by

( A.10 ) $$A_0^T W = -\begin{bmatrix}A_1^T & \cdots & A_p^T\end{bmatrix}$$

This provides a method for obtaining the matrix of causal coefficients, $A$, directly from the auto-covariance matrix $\Lambda$. As with the original sparsest permutation algorithm, we can search over all possible permutations of our $n$ variables, recalculate the auto-covariance, and search for the sparsest coefficient matrix $A$.